\documentclass[aps,superscriptaddress,twocolumn,amsfonts,amsmath,showpacs,floatfix]{revtex4}
\input epsf
\usepackage{epsfig}
\usepackage{graphicx}
\epsfclipon
\def \Dmax {\Delta_{\rm max}}
\def \Dmin {\Delta_{\rm min}}

\begin{document}

\title{Density minimum and liquid-liquid phase transition}

\author{Peter H. Poole} 
\affiliation{Department of Physics, St. Francis Xavier University,
Antigonish, Nova Scotia B2G 2W5, Canada}

\author{Ivan Saika-Voivod} \affiliation{Dipartimento di Fisica and
INFM-CRS-Soft, Universita' di Roma La Sapienza, Piazzale Aldo Moro~2,
I-00185, Roma, Italy} \affiliation{Department of Chemistry, University
of Saskatchewan, Saskatoon, Saskatchewan S7N 5C9, Canada}

\author{Francesco Sciortino} \affiliation{Dipartimento di Fisica and
INFM-CRS-Soft, Universita' di Roma La Sapienza, Piazzale Aldo Moro~2,
I-00185, Roma, Italy}

\begin{abstract}
We present a high-resolution computer simulation study of the equation
of state of ST2 water, evaluating the liquid-state properties at 2718
state points, and precisely locating the liquid-liquid critical point
(LLCP) occurring in this model.  We are thereby able to reveal the
interconnected set of density anomalies, spinodal instabilities and
response function extrema that occur in the vicinity of a LLCP for the
case of a realistic, off-lattice model of a liquid with local
tetrahedral order.  In particular, we unambiguously identify a density
minimum in the liquid state, define its relationship to other
anomalies, and show that it arises due to the approach of the liquid
structure to a defect-free random tetrahedral network of hydrogen
bonds.
\end{abstract}

\date{April 21, 2005}


\maketitle

Water is not only the most abundant liquid on earth, but also a
prototype of many network forming materials.  As for water, important
substances such as silicon and silica exhibit in their liquid phase a
disordered network structure that arises due to highly directional
tetrahedral bonding interactions.  It has long been appreciated that
the development of an open tetrahedral network on cooling is related
to the occurrence of the density maximum observed in these
liquids~\cite{ak76}. In water the density maximum occurs at $277$~K at
ambient pressure $P$.  For temperatures $T$ above the density maximum,
the isobaric expansivity $\alpha_P$ is positive, as for the vast
majority of liquids.  As $T$ decreases through the density maximum,
$\alpha_P$ decreases to zero and then becomes negative; in this region
the liquid expands on cooling.  For all liquids with density maxima,
an open question concerns what happens as $T$ decreases further: Does
$\alpha_P$ remain negative, or eventually return to ``normal''
positive values?  In the latter case, the $T$ at which $\alpha_P$
again becomes positive corresponds to a density minimum.

A density maximum, although well-known in the case of water, occurs in
only a few liquids.  Experimental observations of liquid-state density
minima are almost non-existent; we have found only one report in the
literature, a study of a mixture of Ge and Se~\cite{RT76}.  Yet the
possibility of a density minimum occurring in deeply supercooled water
and similar systems has been discussed in the literature (see
e.g. Ref.~\cite{angell00}), and is supported by several theoretical and
simulation studies.  Density minima have been identified in lattice
models of network-forming fluids~\cite{rd96}, and recent computer
simulations of water~\cite{BGO01,GG03,P04} also present data in which
a liquid-state density minimum, or a trend toward one, is seen or can
be inferred.  Also notable is the recent experimental report of a
density minimum in vitreous silica, although non-equilibrium effects
complicate the interpretation of the results in terms of equilibrium
liquid-state behavior~\cite{silica}.

At the same time, a growing body of evidence suggests that a
liquid-liquid critical point (LLCP) may also be a characteristic
feature of liquids with local tetrahedral
order~\cite{PSES92,MS98,SA03,M04}.  Below such a LLCP, two distinct
liquid phases, the so-called ``low density'' and ``high density''
liquids (respectively LDL and HDL) are separated by a line of first
order phase transitions.  Formal thermodynamic analysis has revealed
general relationships that must be satisfied in a liquid displaying
both density anomalies and the anomalies associated with a
LLCP~\cite{speedy,rd96,dd1,dd2,deben1,deben2}.  Some studies, such as
in Ref.~\cite{silica}, associate multiple density anomalies with a
polyamorphic transition related to a LLCP.  However, an unambiguous
realization of these relationships for a realistic, off-lattice,
equilibrium molecular liquid has not been reported, either
experimentally or in computer simulations.  In particular, the
implications of a liquid-state density minimum merit focused study,
both in terms of its relationship to the LLCP, as well understanding
its origins in terms of the low $T$ thermodynamic and structural
behavior of the liquid.  For example, if a density maximum is followed
at lower $T$ by a density minimum, it means that many, if not all of
the anomalies for which water in particular is famous, ultimately
disappear under sufficient supercooling.  The possibility of such a
return to ``normal'' behavior has important implications for
understanding the properties of bulk supercooled water, the amorphous
ices, as well as composite systems in which the tetrahedral network is
influenced by surfaces or solutes~\cite{BGO01,P04}.

In this Letter, we present a high-resolution computer simulation study
of the equation of state (EOS) of ST2 water~\cite{st2}, a molecular
dynamics model known to exhibit a prominent density maximum and also a
clear example of a LLCP~\cite{PSES92}.  As shown below, we find in
addition that this model liquid displays a density minimum.  We are
thereby able to reveal the thermodynamic context in which this density
minimum occurs, identify the structural origins of the anomaly, and
clarify its relationship to the LLCP.

Our results are based on molecular dynamics simulations of liquid
systems of $N=1728$ molecules.  All simulations are carried out at
constant $N$ and volume $V$, and $T$ is maintained at the desired
value using Berendsen's method~\cite{beren} with a time constant for
the relaxation of $T$ fluctuations of $2$~ps.  Molecular interactions
are cut off at a distance of $0.78$~nm, and the reaction field method
is used to approximate the contribution of dipole-dipole interactions
at longer range.  The time step used is $1$~fs.

Our simulation protocol is as follows: Randomized starting
configurations are prepared at each density $\rho$ to be studied and
are used to initiate runs at $400$~K.  An equilibration and production
phase are carried out, after which the target $T$ is reduced by $5$~K,
and a new equilibration/production cycle is initiated using the last
configuration of the previous run.  In this way, the properties of
successive state points along the specified isochore (i.e. at constant
$\rho$) are evaluated with decreasing $T$.  Both the equilibration and
production phases at each $T$ are carried out for the time required
for the mean squared displacement (MSD) to reach $1$~nm$^2$, or for
$100$~ps, whichever is larger.  Note that a MSD of $1$~nm$^2$
corresponds to an average displacement of about 10 times the distance
between nearest neighbour molecules in liquid water at
$\rho=1.0$~g/cm$^3$. For our lowest $T$ state points, our run-time
criterion leads to simulations of on the order of $100$~ns.  Using the
above protocol, we evaluate the properties of the liquid at $2718$
state points: at densities from $\rho=0.70$ to $1.50$~g/cm$^3$ in
steps of $0.01$~g/cm$^3$; and for $T$ from $400$~K down to at least
$255$~K, in $5$~K steps.  Along some isochores, a $T$ as low as
$200$~K is reached.  The sum of the computational work reported here
corresponds to approximately 30 cpu-years on a $3.0$~GHz Intel Xeon
processor.

\begin{figure}
\includegraphics*[width=0.9\linewidth]{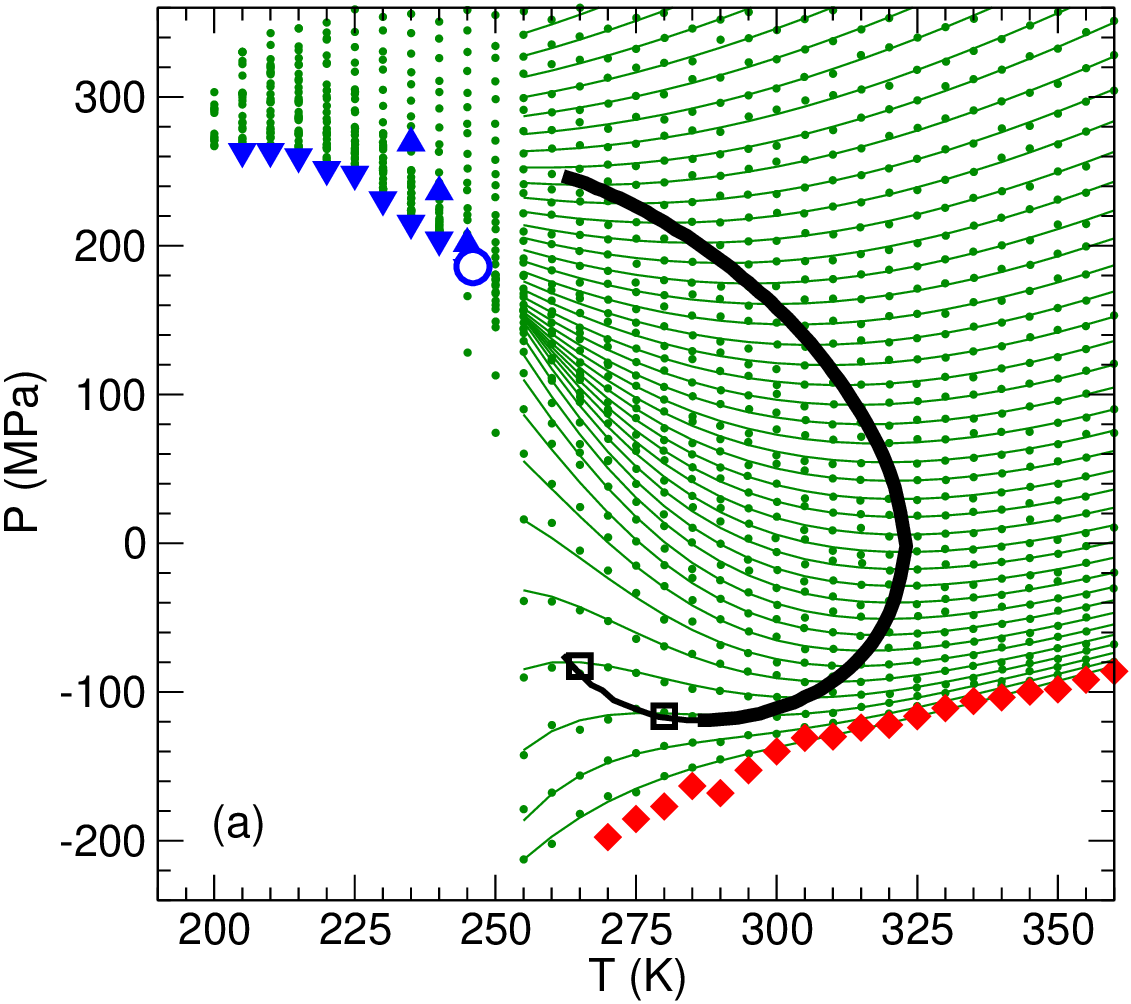}
\includegraphics*[width=0.9\linewidth]{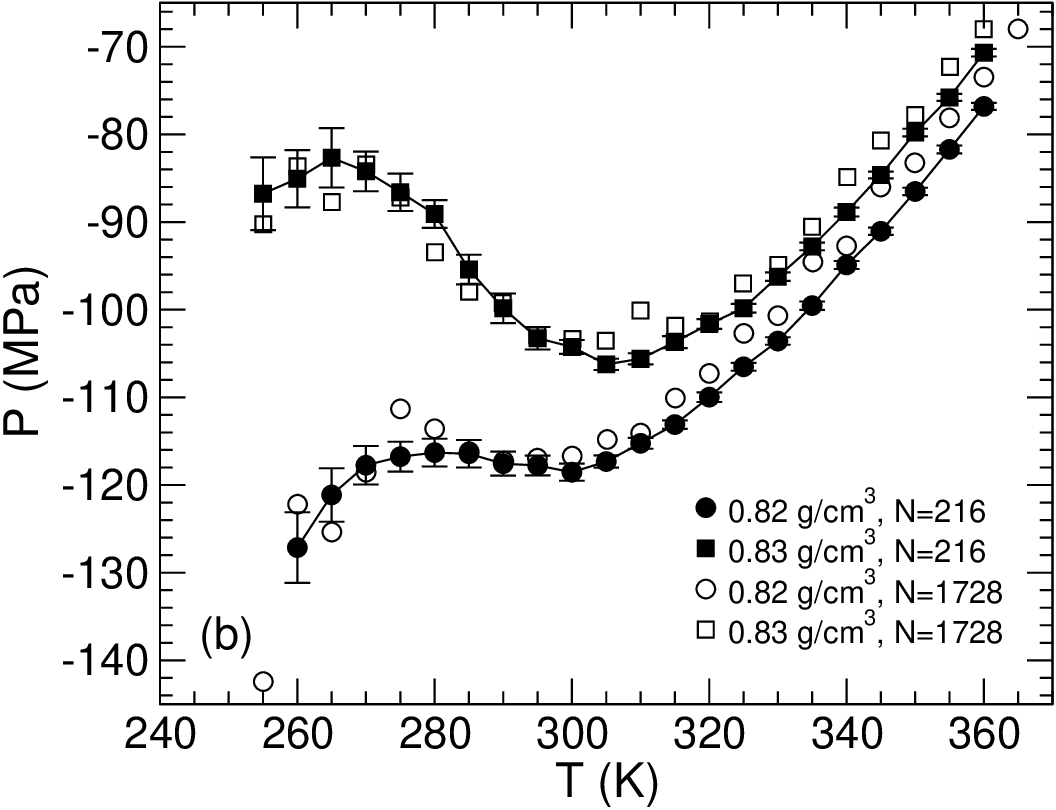}
\caption{(color online) (a) Phase diagram of the $N=1728$ ST2 liquid.
State points simulated are shown as small dots.  Isochores (thin lines
through dots) obtained from our fitted EOS are shown from
$\rho=0.80$~g/cm$^3$ (bottom-most line) and greater in steps of
0.01~g/cm$^3$.  Also shown is the estimated location of the liquid-gas
spinodal (diamonds), the LDL spinodal (up-triangles), the HDL spinodal
(down-triangles), and the LLCP (circle).  $\Dmax$ (thick line)
transforms into $\Dmin$ (thin line) at low $P$.  Squares locate points
on $\Dmin$ obtained from (b).  (b) Variation of $P$ with $T$ along the
$\rho=0.82$ and $0.83$~g/cm$^3$ isochores for both the $N=1728$ and
$N=216$ systems.  For the $N=216$ data, error bars are drawn at twice
the standard deviation of the means of $P$ as evaluated from each of
the 40 runs conducted at each state point.}
\label{data}
\end{figure}

\begin{figure}
\includegraphics*[width=0.9\linewidth]{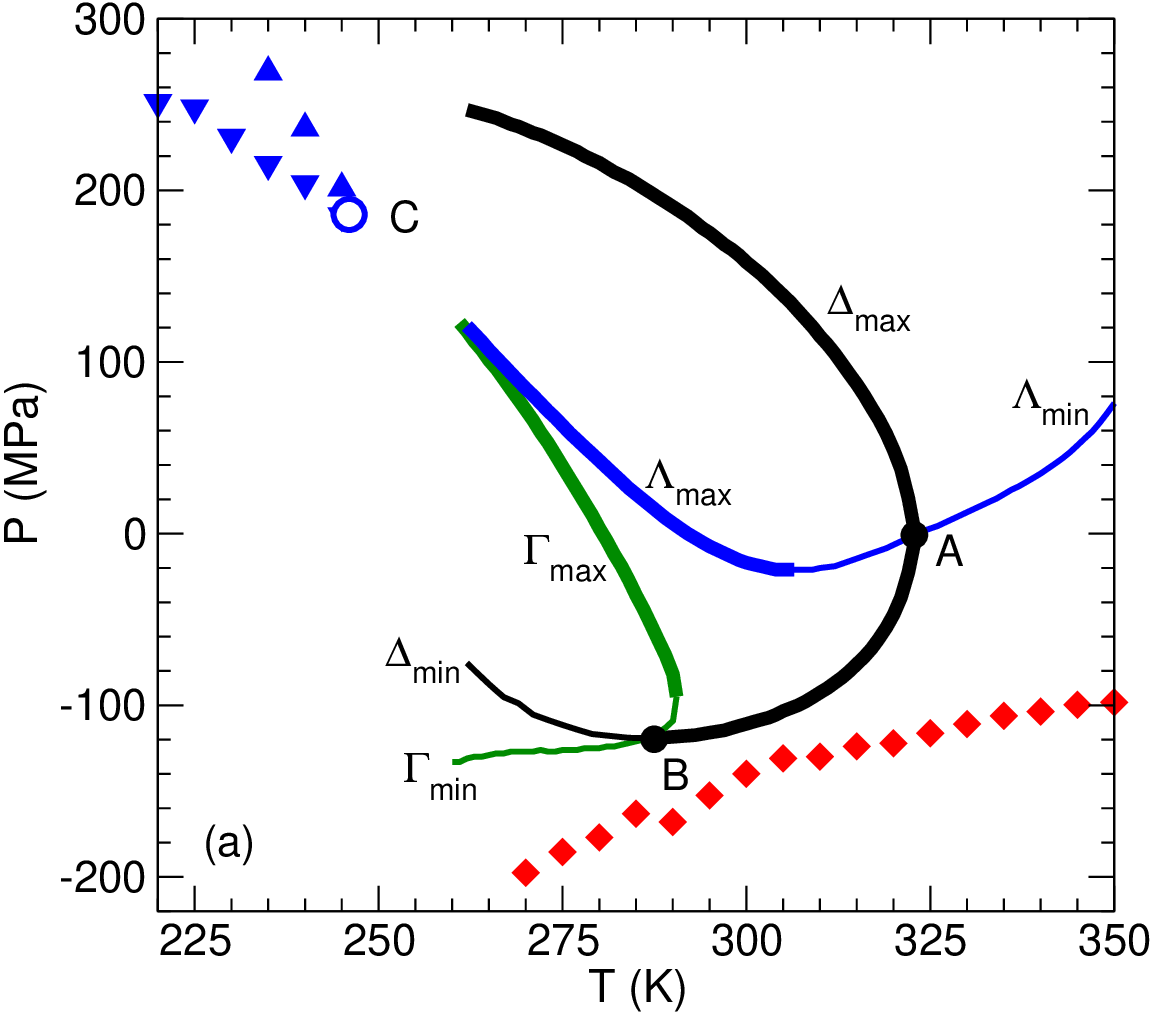}
\includegraphics*[width=0.9\linewidth]{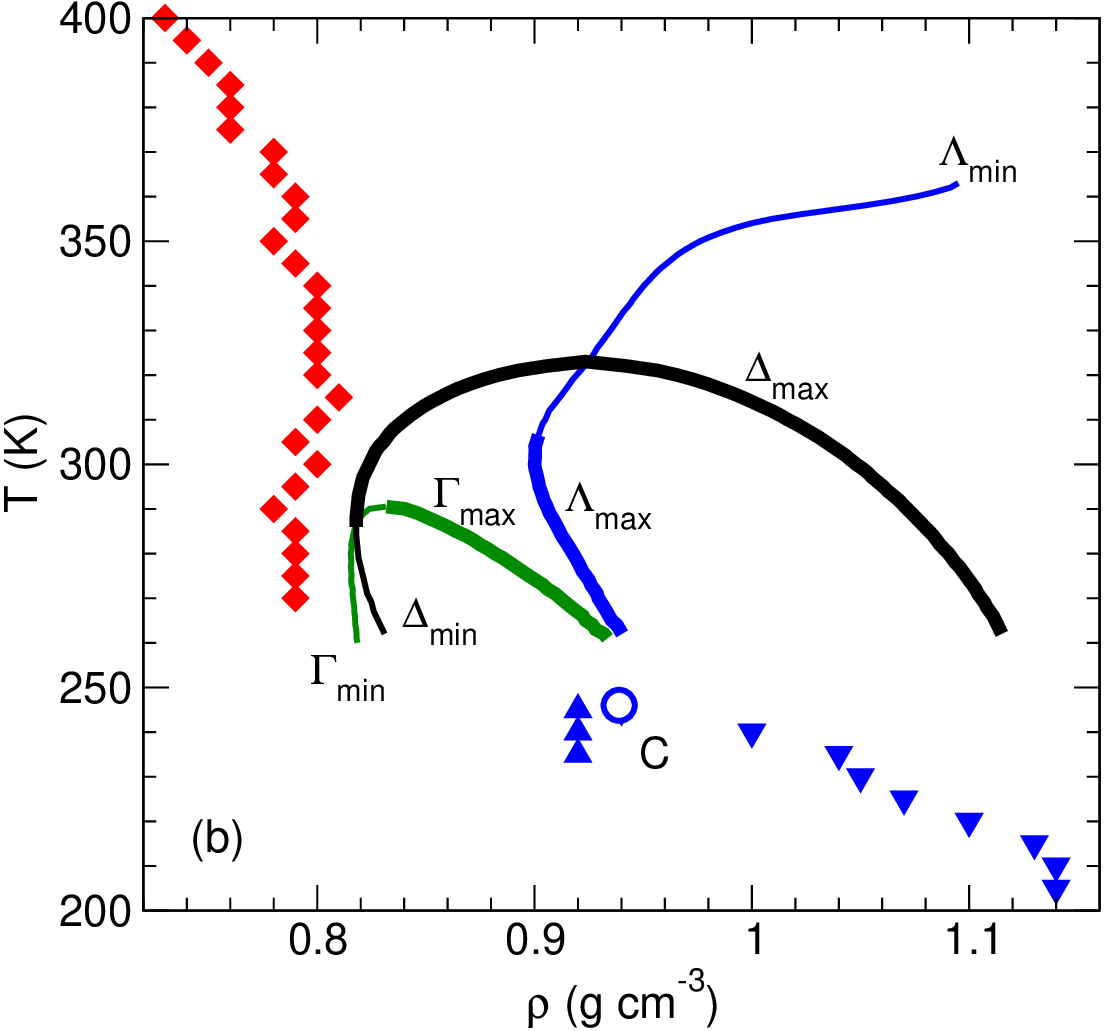}
\caption{(color online) Location of the LLCP ($C$), spinodals
(triangles and diamonds), density extrema ($\Delta$), and response
function extrema ($\Lambda$ and $\Gamma$) in (a) the $T$-$P$ and (b)
the $\rho$-$T$ planes.  The symbols used for the various spinodals are
the same as in Fig.~\protect{\ref{data}}.  Thick lines indicate
maxima, and thin lines locate minima.}
\label{fit}
\end{figure}

The resulting EOS data for $P(\rho,T)$ is shown in Fig.~\ref{data}(a).
Early support for the possibility of a LLCP in water was based on the
shape of the ST2 EOS reported in Ref.~\cite{PSES92}.  These features
of the ST2 model are seen here in much greater detail.  The LLCP is
readily located, as well as the metastability limits (spinodals) of
both the LDL and HDL phases that meet at the critical point.  We also
identify an upper bound on the location of the liquid-gas spinodal by
locating the cavitation limit of the liquid at the extreme low $\rho$
range at negative $P$.

Fig.~\ref{data}(a) shows most of the individual state points
simulated, as well as isochores obtained from the fit of an analytic
EOS to the data set~\cite{method1}.  Points on the locus of density
maxima $\Dmax$ are coincident with the minima of these isochores.  As
$\rho$ decreases, $\Dmax$ follows a retracing path in the $T$-$P$
plane to $\rho=0.82$~g/cm$^3$, but notably, the isochores for
$\rho=0.81$ and $0.80$~g/cm$^3$ do not exhibit minima.  Rigorous
thermodynamic arguments have shown that $\Dmax$ cannot end at finite
$T$ in isolation from other anomalies~\cite{speedy,dd1,dd2}.  It must
either terminate on a spinodal locus, or $\Dmax$ itself must pass
through an extremum in the $T$-$P$ plane, at which point it becomes a
line of density minima, $\Dmin$.  We find that the latter possibility
occurs, as shown by both the shape of $\Dmax$ and the EOS isochores:
low $T$ maxima in the $\rho=0.82$ and $0.83$~g/cm$^3$ isochores
correspond to points on $\Dmin$.

To further confirm the existence of $\Dmin$, we conduct a new set of
simulations focusing on the $\rho=0.82$ and $0.83$~g/cm$^3$ isochores,
with the aim of reducing the statistical errors.  These simulations
employ a smaller system size, $N=216$, but we average over an ensemble
of 40 independent runs for each state point~\cite{method2}. The
variation of $P$ with $T$ along these two isochores
[Fig.~\ref{data}(b)] confirms that both a density maximum and a
density minimum occur.  The locations of the density minima so
identified are shown as squares in Fig.~\ref{data}(a), and are
consistent with the predictions of the data set obtained using $1728$
molecules.

All along the locus $\Delta$ (the union of $\Dmax$ and $\Dmin$) the
condition $(\partial \rho/\partial T)_P=0$ is satisfied.  Formal
thermodynamic analysis has shown that changes of sign of the slope of
$\Delta$ in the $T$-$P$ plane are associated with intersections with
certain response function extrema.  Ref.~\cite{deben1} shows that the
point $A$ in Fig.~\ref{fit}(a), where $\Delta$ has infinite slope, is
coincident with a point on a locus $\Lambda$ along which $(\partial
K_T/\partial T)_P=0$, where $K_T$ is the isothermal compressibility.
By evaluating the appropriate derivative of our fitted EOS, we confirm
that this requirement is satisfied.  We also identify a new
thermodynamic relation, derived by analogy to those in
Refs.~\cite{deben1} and \cite{deben2}, that states that the point $B$
in Fig.~\ref{fit}(a), where $\Delta$ has zero slope, is coincident
with a point on a locus $\Gamma$ along which $(\partial C_P/\partial
P)_T=0$, where $C_P$ is the isobaric specific heat.  This requirement
is also satisfied by our fitted EOS. Fig.~\ref{fit}(b) shows the
location of the features of (a) in the $\rho$-$T$ plane.

\begin{figure}
\includegraphics*[width=0.9\linewidth]{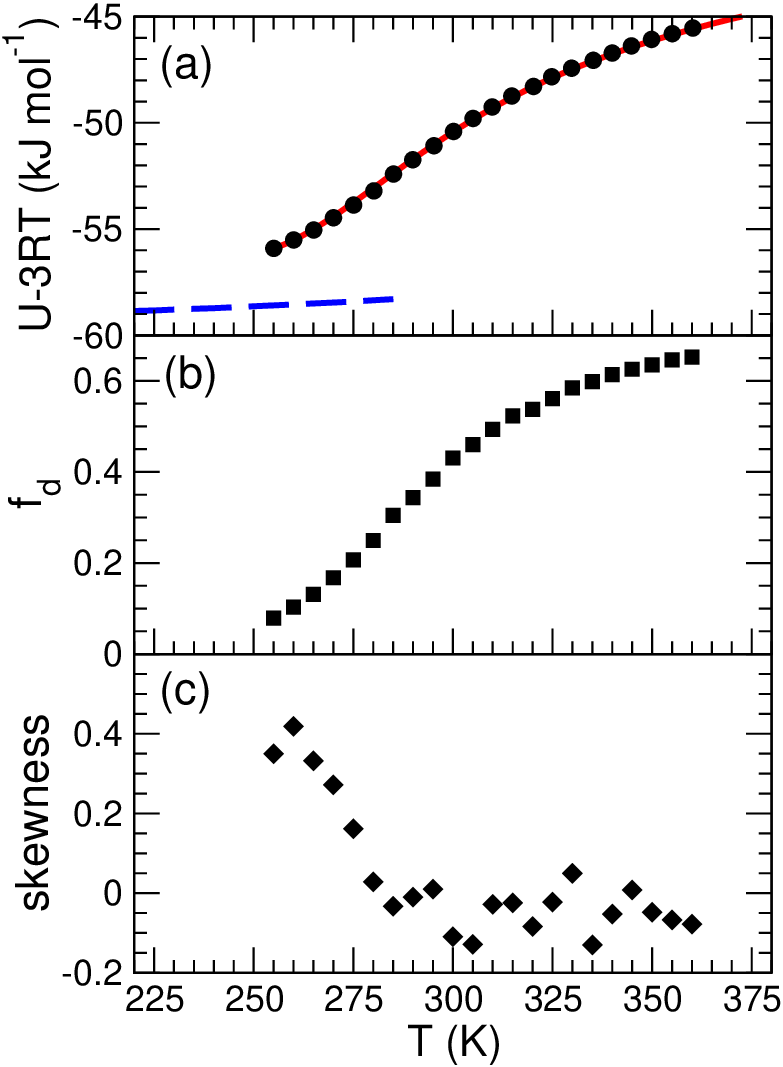}
\caption{(color online) Properties of the liquid along the
$\rho=0.83$~g/cm$^3$ isochore as found from the $N=216$ simulations.
(a) Variation of $U-3RT$ with $T$ for the liquid (circles), compared
with $U-3RT$ for ice~I$_h$ (line).  $R$ is the gas constant.  The
dashed line shows $U-3RT$ from the fitted EOS.  $U$ for ice~I$_h$ is
found from simulations of a proton-disordered configuration of $N=432$
ST2 molecules at $\rho=0.83$~g/cm$^3$. (b) The fraction of network
defects $f_d$ as a function of $T$.  $f_d$ is estimated by evaluating
the equilibrium average fraction of molecules not having exactly four
nearest neighbour O atoms within $0.35$~nm, the distance of the first
minimum in the O-O radial distribution function.  (c) The skewness of
the $P(U)$ distributions plotted in Fig.~\protect{\ref{dist}}.  The
skewness is evaluated as $(1/n)\sum_{i=1}^n [(U_i-{\bar
U})/\sigma]^3$, where $n$ is the number of $U_i$ values sampled, $\bar
U$ is the average value of $U$, and $\sigma$ is the standard deviation
of $P(U)$.}
\label{huge}
\end{figure}

At the same time, the curves $\Lambda$ and $\Gamma$ are formally
related to the LLCP.  At a LLCP, denoted $C$, isotherms of $C_P$ and
isobars of $K_T$ will both exhibit a divergence.  For $T>T_C$,
$\Lambda$ and $\Gamma$ radiate outward from $C$ as lines of maxima
that are the supercritical extension of the divergence occurring as
$T\to T_C$.  In the present data, these maxima transform into minima,
and then intersect the line $\Delta$, controlling the sign of its
slope.  While the existence of these response function extrema is not
a sufficient condition for there being a LLCP at lower $T$, it is a
necessary condition~\cite{deben1}.  The interconnected pattern of
density anomalies and response function extrema shown here therefore
illustrates how the discovery of such features in experimental data
can provide a rational procedure for locating a LLCP at lower $T$.

Examination of the thermodynamics and structure of the liquid in the
vicinity of the density minimum reveals why it occurs.  As shown in
Fig.~\ref{huge}, both the potential energy $U$, and the fraction of
network defects, $f_d$, are at the lowest $T$ behaving more and more
as they would for a system approaching a random tetrahedral network
(RTN) in which all nearest-neighbour intermolecular bonds are
satisfied: the $T$ dependence of $U$ approaches an ice-like form, and
$f_d$ approaches zero.  The occurrence of inflections in both these
curves shows that the region of most rapid change has passed, and that
the low $T$ region in which the density minimum occurs corresponds to
the final asymptotic consolidation of the liquid structure towards
that of a perfect, defect-free RTN.  It is plausible that $\alpha_P$
should be positive (i.e. normal) for a defect-free RTN;
e.g. $\alpha_P>0$ for crystalline ice I$_h$ (an {\it ordered}
tetrahedral network) in this $T$ range~\cite{japan}.  In this light,
the density minimum simply reflects the fact that the process of RTN
formation is coming to an end, and so the anomalous $T$ dependence of
thermodynamic properties returns to normal.

\begin{figure}[t]
\includegraphics*[width=0.9\linewidth]{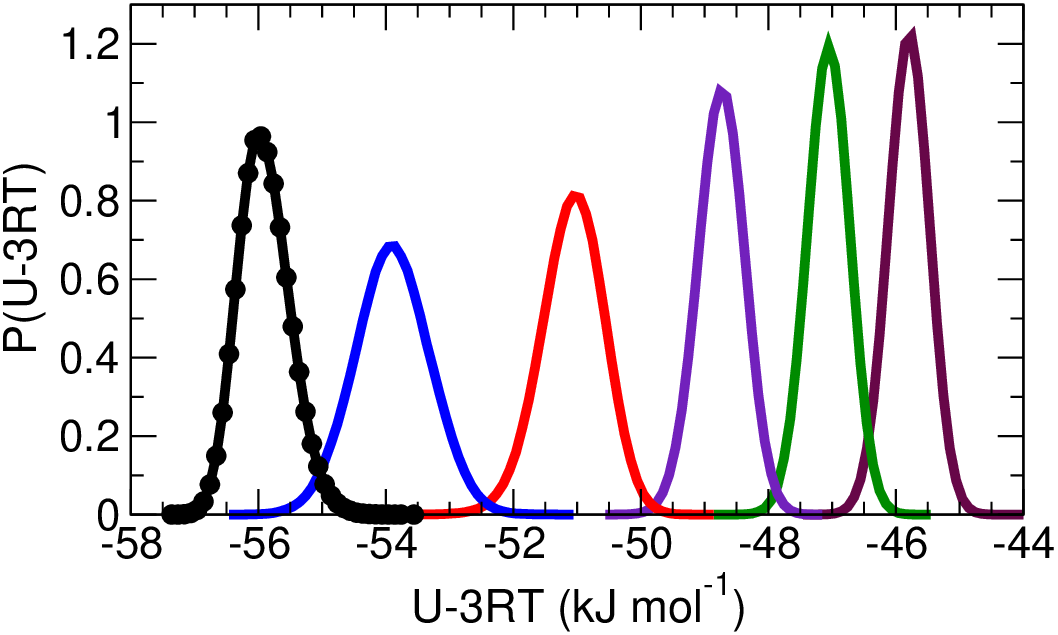}
\caption{(color online) Probability distributions of $U$ along the
$\rho=0.83$~g/cm$^3$ isochore as found from the $N=216$ simulations.
From left to right are shown the distributions for $T=255$, 275, 295,
315, 335 and 355~$K$. $R$ is the gas constant.}
\label{dist}
\end{figure}

Another indicator that the liquid is approaching the RTN state can be
seen in the distribution of values of $U$, the potential energy,
sampled in our simulations (Fig.~\ref{dist}).  For $T$ less than the
inflection in $U$, the distribution $P(U)$ narrows, and also loses its
symmetrical shape and becomes skewed, as quantified in
Fig.~\ref{huge}(c).  This narrowing and skewing suggests that there is
a lower bound on the value of $U$, specifically $U$ for the RTN, and
that the system is entering the $T$ regime where the existence of this
lower bound influences the average properties~\cite{heuer}.

The location and slope of the $\Dmin$ line in Fig.~\ref{data} raises
the possibility that a density minimum occurs in ST2 water at positive
$P$ at sufficiently low $T$.  This, and the generic nature of the
mechanism underlying the density minimum found here, demonstrates the
potential for the occurrence of density minima in the low $T$, $P>0$
behaviour of any tetrahedral liquid having a density maximum.  More
broadly, our results add to and illuminate the complex behaviour of
tetrahedral liquids: The emergence of a RTN structure in these liquids
can be understood as the source of their anomalies. Yet the eventual
outcome of this RTN-forming process is to bring the liquid closer and
closer to a set of near-perfect RTN states that lie at the bottom of
the liquid's potential energy landscape~\cite{SPS01,heuer,moreno}.
Our simulations show that such a ``liquid at the bottom of the
landscape'' represents a distinct regime of behaviour: one that
exhibits a number of exotic properties not observed at higher $T$
(e.g. a crystal-like heat capacity), and also one in which
thermodynamic anomalies, so often associated with networking-forming
liquids, disappear.

We thank the StFX hpcLAB and G. Lukeman for computing resources and
support.  PHP and ISV acknowledge support from the AIF, the CFI, the
CRC Program, and NSERC.  FS acknowledges support from Miur-Firb.


\begin{thebibliography}{999}
\bibitem{method1} We create a fitted EOS model of the data for
$P(V,T)$ and $E(V,T)$ for $255$~K~$\le T \le 400$~K and
$0.80$~g/cm$^3$~$\le \rho \le 1.30$~g/cm$^3$, where $E$ is the total
energy.  This encompasses 1530 state points.  To model both the
$P(V,T)$ and $E(V,T)$ data, we carry out a least-squares fit to a
bivariate polynomial of the form $\sum_{i=0}^{7} \sum_{j=0}^{6} a_{ij}
c_i(V) c_j(T)$ where $c_i(V)$ is the Chebyshev polynomial of the first
kind of degree $i$ as a function of $V$, $c_j(T)$ is similarly
defined, the $a_{ij}$'s are the fit parameters.  Extrema of $\rho$ and
$K_T$ are found from the fit to the $P(V,T)$ data.  Extrema of $C_P$
are found by isobarically differentiating the enthalpy $H(P,T)=E+PV$.
\bibitem{method2} In the case of the $N=216$ system, we carry out the
same protocol as in the $N=1728$ system, 40 times per isochore for the
two cases $\rho=0.82$ and $0.83$~g/cm$^3$, each time using a
randomized starting configuration to initiate the first run at
$T=360$~K.  Also, the minimum production run time is increased to
$500$~ps for all state points.
\end{thebibliography}
\end{document}